\documentclass[aps,prd,preprint,superscriptaddress,floatfix]{revtex4-2}
\usepackage{amsmath,amssymb}
\usepackage{graphicx}
\usepackage{url}
\usepackage{bm}
\usepackage{amsthm}
\makeatletter
\def\frontmatter@abstract@produce{%
  \par
  \addvspace{\frontmatter@preabstractspace}%
  \begingroup
    \dimen@\baselineskip
    \setbox\z@\vtop{\unvcopy\absbox}%
    \advance\dimen@-\ht\z@\advance\dimen@-\prevdepth
    \@ifdim{\dimen@>\z@}{\vskip\dimen@}{}%
  \endgroup
  \begingroup
    \prep@absbox
    \unvbox\absbox
    \post@absbox
  \endgroup
  \@ifx{\@empty\mini@notes}{}{\mini@notes\par}%
  \addvspace\frontmatter@postabstractspace
}
\makeatother

\begin{document}

\title{Data-Driven, Geometry-Aware Optimal-Transport Calibration of Flavor Taggers}
\author{Yeonjoon Kim}
\email{kyj519@snu.ac.kr}
\affiliation{Department of Physics and Astronomy, Seoul National University, 1 Gwanak-ro, Gwanak-gu, Seoul 08826, Korea}
\author{Un-Ki Yang}
\email[Corresponding author: ]{ukyang@snu.ac.kr}
\affiliation{Department of Physics and Astronomy, Seoul National University, 1 Gwanak-ro, Gwanak-gu, Seoul 08826, Korea}

\begin{abstract}
Flavor-tagging calibrations are often provided either as scale factors measured at a finite set of working points or as binned corrections to a chosen one-dimensional discriminant. However, this approach falls short of providing continuous, event-level calibration across the full multicomponent outputs of modern taggers. This limitation leads to information loss in analyses that demand high-performance flavor tagging, effectively restricting the use to a single definition among several viable tagger variables that can be defined by specific proportions of the multi-head output.

In this work, we propose a geometry-aware framework that formulates flavor-tagger calibration as an optimal transport problem on the probability simplex. The transport maps are parameterized and trained in the isometric log-ratio (ILR)

coordinate system. Because the quadratic Euclidean cost of Brenier transport in this coordinate system is equivalent to the Aitchison distance on the simplex, the learned map induces a controlled minimal deformation with respect to the Aitchison metric of the underlying flavor probability structure.

Furthermore, we extract flavor-conditional target distributions directly from control-region data using an expectation-maximization (EM) technique that simultaneously fits multiple control regions, models each flavor component with a normalizing flow, and estimates the regional mixture fractions. In the simulation study presented here, this data-extraction step is validated using pseudo-data control regions. Monte Carlo simulation is used for the source distribution, component pretraining, and control-region design. The extracted targets are subsequently used to learn flavor-factorized transport maps.
 Because the joint estimation of mixture fractions and flexible component densities admits weakly constrained directions, we further introduce a linearized feedback-operator analysis that propagates the fitted composition covariance into the extracted component densities, separating data-constrained modes from those dominated by the composition prior.

The simulation-based closure study demonstrates improved closure in dedicated control regions and in independent validation mixtures.
\end{abstract}


\maketitle

\section{Introduction}
High-precision measurements and searches at the LHC depend critically on robust calibration of flavor-tagging observables. The standard strategy absorbs data--simulation mismodeling through scale factors (SFs). Although this methodology is mature and extensively validated~\cite{cms_btag_run2,atlas_btag_run2,cms_cjet_calib}, it has several structural limitations.

First, SF extraction is typically tied to predefined working points, which requires a binned treatment of the tagger response. Since the underlying response is continuous, this discretization inevitably discards information. Second, one must define a specific scalar score before calibration. Modern taggers provide multiple outputs associated with class probabilities~\cite{gn2} (e.g. $b$, $c$, and light flavors), and analyses usually construct task-specific combinations such as
\begin{equation}
\frac{p_b}{p_b+p_c}, \qquad \frac{p_c}{p_c+p_\ell}, \qquad \frac{p_b}{p_b+\kappa_c p_c}.
\end{equation}
As a consequence, the calibration product is often tied to a specific score definition.
In addition, the optimal trade-off between signal-flavor efficiency (typically $b$ or $c$ jets) and background mistag rejection (typically light jets) is strongly analysis-dependent. Different analyses therefore require different working points or even different score parameterizations, which in turn implies different SF measurements. In the limiting case, each analysis benefits from its own dedicated calibration campaign, as also noted in previous OT-based calibration studies~\cite{atlas_ot}.

To address these issues, continuous-score calibration strategies have been proposed, including CMS iterative-fit style calibrations for heavy-flavor tagger scale-factor extraction~\cite{cms_cjet_calib} and OT-based approaches in ATLAS~\cite{atlas_ot}. The iterative-fit approach still behaves effectively as a fine-binned SF calibration and remains tied to a chosen score definition. OT-based approaches are conceptually closer to the present work, but existing OT-based flavor-calibration studies have so far demonstrated the method in restricted calibration settings, such as $b$-tagging applications~\cite{atlas_ot}, and generalization to broader score definitions requires careful design of the calibration space and transport objective.
Consequently, this work aims for a calibration of the full probability manifold of the flavor tagger, utilizing all continuous information of the tagger response while simultaneously calibrating various derived scores resulting from operations among individual tagger components at once.

The EM procedure described below yields normalizing-flow models for both the MC component densities $q_f^{\mathrm{MC}}$ (from supervised pretraining) and the data component densities $q_f^{\mathrm{data}}$ (from EM applied to control-region data). One might therefore ask whether an explicit transport map is even necessary: given both NF models, one could in principle form the pointwise density ratio $q_f^{\mathrm{data}}(x)/q_f^{\mathrm{MC}}(x)$ and use it as a continuous, event-level scale factor. This ratio is well defined wherever $q_f^{\mathrm{MC}}(x)>0$, and for binned working points its bin-integrated value reproduces the conventional scale factor.

For a continuous score in the two-dimensional ILR space, however, the density-ratio approach becomes significantly less robust. In the binned working-point case, each bin integrates the density over a finite volume, so localized regions of low MC support are averaged into finite event counts and the resulting correction remains well behaved. For a continuous score, this bin-level regularization is absent. In regions where the MC density becomes very small relative to the data, the pointwise ratio can become numerically unstable and highly sensitive to finite-sample fluctuations or modelling imperfections.

A transport map provides a more robust alternative in this setting. Rather than relying on a pointwise density ratio, it seeks a deterministic rearrangement $T_f$ such that $T_{f\#}\mu_f^{\mathrm{MC}} \approx \mu_f^{\mathrm{data}}$, moving probability mass coherently across the ILR space. Its practical advantage is that, in the finite-sample regime relevant here, it remains substantially more stable than pointwise density-ratio reweighting when the MC density is low but nonzero and the estimated ratio would otherwise become ill-conditioned. For the continuous support-mismatch effects considered in this work, the transport-map formulation is therefore the more robust choice.

A second issue, less often discussed in the calibration literature, is that data-driven extraction of flavor-conditional tagger responses is itself a non-convex inference problem. When mixture fractions and flexible component densities are estimated jointly, the optimization can admit nearly flat directions along which composition shifts and density deformations partially cancel. A flexible model that reproduces the observed control-region distributions does not, by itself, guarantee that the extracted flavor-conditional response is uniquely determined by the data. Any calibration framework that relies on data-driven target extraction therefore needs an explicit treatment of identifiability and a tractable prescription for propagating composition uncertainty into the calibrated response.

The main contributions of this work are: (i)~formulating multiclass tagger calibration as an optimal transport problem on the probability simplex, with ILR/Aitchison geometry ensuring that the transport cost is aligned with the intrinsic structure of compositional probability vectors; (ii)~extracting flavor-conditional data targets from multiple enriched control regions via an expectation-maximization procedure with normalizing-flow components~\cite{rezende_flow,papamakarios_flow,rqs}; (iii)~validating the resulting flavor-factorized Brenier OT maps in a simulation closure study, demonstrating closure on derived discriminants that were not used during training; and (iv)~introducing a tractable uncertainty prescription based on the fitted mixture-composition covariance and a linearized feedback operator, which separates data-constrained directions from those dominated by the composition prior.
The central methodological point of this work is that flavor-tagging scores are simplex-constrained probability vectors. Therefore, coordinate choices and distance definitions are not technical details; they define what is actually minimized during calibration. We adopt an ILR embedding and solve quadratic OT in that space. This gives a map that is both mathematically controlled and physically interpretable in probability space.

A key architectural assumption underlying the EM extraction is transferability: the flavor-conditional tagger response $q_f(x)$ is assumed to be the same across control regions after fixing the relevant kinematic variables. This assumption is automatically satisfied in the simulation closure study presented here, but its validation in real collision data---via fine phase-space binning or other approaches---is an experimental requirement that lies outside the scope of this paper.

\section{Problem formulation}
\label{sec:problem}

In this section, we formalize the map-based calibration objective and define the transport geometry used throughout the analysis.

\subsection{Flavor tagger outputs}
The output of a flavor tagger is generally written as
\begin{equation}
\bm{p} = (\{p_f\}), \qquad p_f>0, \qquad \sum_{f}p_f = 1,
\end{equation}
where $f$ denotes the flavor categories resolved by the tagger. In practical analyses, quarks and antiquarks are typically not separated, and jets initiated by $u,d,s$ quarks and gluons are grouped into a single light-flavor category. We therefore use $f\in\{b,c,\ell\}$ in this paper, so that
\begin{equation}
\bm{p} \in \mathcal{S}^2,
\end{equation}
that is, the output vector is assumed to lie on the two-dimensional probability simplex.

\subsection{Brenier OT}
Let $\mu_{\mathrm{MC}}$ and $\mu_{\mathrm{data}}$ denote the score distributions
in Monte Carlo and data. A deterministic calibration map $T$ is defined by
\begin{equation}
    T_{\#}\mu_{\mathrm{MC}}=\mu_{\mathrm{data}},
\end{equation}
where $T_{\#}$ denotes the pushforward operator.

The objective of map-based calibration is to find a map $T$ that transports the MC distribution to the data distribution while minimizing a physically meaningful deformation cost:
\begin{equation}
T^{\star} = \arg\min_{T_{\#}\mu=\nu} \int c\!\left(\bm{p},T(\bm{p})\right)\,d\mu(\bm{p}),
\end{equation}
where $\mu$ and $\nu$ denote the source and target probability measures, respectively.

A practical caveat is that the Monge formulation does not, by itself, guarantee that a neural OT training procedure will converge to a unique map in finite-sample implementations. As discussed by Makkuva \emph{et al.}~\cite{makkuva}, neural-network-based OT training can converge to different transport solutions depending on initialization and optimization details. We therefore follow the Brenier-OT framework also adopted in previous calibration studies~\cite{atlas_ot}, where the map is modeled as the gradient of a convex potential and trained with a quadratic Euclidean cost~\cite{brenier}. 

Our main distinction is not the use of the Brenier framework itself, but the choice of calibration space. While previous approaches have considered other score parameterizations, in this work we perform the transport in ILR coordinates, so that the Euclidean cost used by the OT objective corresponds to the Aitchison geometry of the probability simplex. This choice is intended to align the calibration objective with the compositional structure of multiclass tagger outputs and to reduce arbitrariness in how distances are measured between probability vectors.
\subsection{Aitchison geometry and ILR coordinates}
In the present context, Brenier's theorem is naturally formulated in an unbounded Euclidean space with quadratic cost, whereas flavor-tagging outputs live on the simplex rather than in $\mathbb{R}^n$. A suitable variable transformation is therefore required.

A straightforward approach, consistent with previous ATLAS implementations~\cite{atlas_ot}, is to apply component-wise logit coordinates and minimize Euclidean displacement in that transformed space. However, this parameterization is over-complete relative to the intrinsic degrees of freedom of the simplex and can induce leakage when transported points are mapped back to physical probabilities. Specifically, a probability vector with $D$ components has only $D-1$ independent degrees of freedom because of the closure constraint $\sum_i p_i=1$, and therefore lies on $\mathcal{S}^{D-1}$. A component-wise logit map embeds this manifold into $\mathbb{R}^D$, so physically admissible points occupy only a $(D-1)$-dimensional submanifold of the ambient space. If the transport map is learned in the full $D$-dimensional space without enforcing this structure, optimization can move samples away from the admissible submanifold, producing unphysical coordinates (``leakage'') during calibration. The redundant dimension also increases statistical and optimization burden in finite samples. More importantly, Euclidean distance in naive logit coordinates is not the canonical metric for simplex-constrained probability vectors, so the optimization objective is not directly aligned with simplex-preserving calibration.

For simplex-constrained probability vectors, the natural metric is the Aitchison distance~\cite{aitchison},
\begin{equation}
d_A(x,y)^2 = \frac{1}{2D}\sum_{i=1}^{D}\sum_{j=1}^{D}
\left(\ln\frac{x_i}{x_j}-\ln\frac{y_i}{y_j}\right)^2,
\end{equation}
which is defined through log-ratios and therefore captures relative changes among class probabilities.

Since the Brenier formulation optimizes Euclidean transport cost, we require a transform that maps Aitchison geometry into a Euclidean one.

We use an ILR embedding~\cite{egozcue_ilr} $z=\mathrm{ILR}(\bm{p})\in\mathbb{R}^{D-1}$; for $D=3$,
\begin{equation}
z=
\begin{pmatrix}
\frac{1}{\sqrt{2}}\ln\left(\frac{p_b}{p_c}\right)\\[3pt]
\sqrt{\frac{2}{3}}\ln\left(\frac{\sqrt{p_b p_c}}{p_\ell}\right)
\end{pmatrix},
\end{equation}
or equivalently via the Helmert sub-matrix $H$,
\begin{equation}
\mathbf{z} = H \log \mathbf{p},
\qquad
H=
\begin{pmatrix}
\tfrac{1}{\sqrt{2}} & -\tfrac{1}{\sqrt{2}} & 0\\[4pt]
\tfrac{1}{\sqrt{6}} & \tfrac{1}{\sqrt{6}} & -\tfrac{2}{\sqrt{6}}
\end{pmatrix}.
\end{equation}
Euclidean distance in ILR space equals the Aitchison distance~\cite{aitchison,egozcue_ilr}; for $D=3$,
\begin{equation}
    d_A(p,q)^2=\frac{1}{6}\sum_{i,j\in \mathcal{F}}
    \left(\ln\frac{p_i}{p_j}-\ln\frac{q_i}{q_j}\right)^2.
\end{equation}
Therefore, quadratic OT in ILR space is directly interpretable as minimum-distortion transport in probability space.

\paragraph{Ratio structure is correctly preserved.}
Euclidean distance in raw probability space does not correctly reflect the probabilistic nature of tagger outputs. Consider two pairs of probability vectors:
\begin{align}
    &p=(0.94,0.04,0.02),\quad q=(0.94,0.02,0.04),\nonumber\\
    &p=(0.40,0.40,0.20),\quad q=(0.40,0.20,0.40).
\end{align}
In both cases the sub-composition score $p_c/(p_c+p_\ell)$ changes from $2/3$ to $1/3$, so calibration should assign equal transport cost to both pairs, especially if the goal is to calibrate all such sub-composition scores consistently. In raw probability space the squared distances are $0.0008$ and $0.08$---a factor of one hundred apart. In ILR space the Aitchison distance gives $\approx 0.98$ in both cases. This is a concrete illustration of how ILR correctly encodes relative-ratio information and reduces distortion.

\paragraph{Prior reweighting corresponds to a translation.}
The ILR embedding has a further practical advantage: reweighting the tagger prior by component factors $(\kappa_b,\kappa_c,\kappa_\ell)$---i.e.\ forming $p'_f=\kappa_f p_f/\sum_j \kappa_j p_j$---corresponds to an additive translation in ILR coordinates. For example, the $\kappa$-parameterized score $p_b/(p_b+\kappa p_c)$ is obtained from the $\kappa=1$ calibration by the ILR translation
\begin{equation}
(z_1,z_2)\;\mapsto\;
\left(
z_1-\frac{\ln\kappa}{\sqrt{2}},\;
z_2+\frac{\ln\kappa}{\sqrt{6}}
\right),
\end{equation}
which can be verified by applying the inverse map $\mathbf{p}=\exp(H^\top\mathbf{z})/\mathbf{1}^\top\exp(H^\top\mathbf{z})$. Consequently, a single ILR-space calibration simultaneously covers all ratio-score parameterizations with arbitrary prior reweighting, without retraining.

\section{Transport map construction in ILR space}
\label{sec:map}

\subsection{Data-based EM target construction}

The response of a flavor tagger is intrinsically flavor-dependent, and so is the pattern of data--MC discrepancy. For instance, in a $b$-enriched region (typically characterized by $p_b \approx 1$), the probability for true $b$ jets to populate that region is often overestimated in simulation, corresponding to a $b$-tag efficiency scale factor below unity. By contrast, the probability for light-flavor jets to enter the same region is often underestimated, corresponding to a mistag scale factor above unity. This motivates a flavor-resolved treatment of the discrepancy, in close analogy with conventional flavor-tagging calibration, and the same logic applies when learning optimal-transport (OT) maps.

A seemingly simple alternative would be to learn a single conditional transport map with the flavor label provided as an auxiliary input. However, under a fixed mixed training sample, this amounts to solving multiple transport problems through a single aggregated objective. The optimizer is then driven by a weighted sum of heterogeneous transport costs, without a first-principles argument that the combined objective should correspond to the physically correct decomposition of the discrepancy. In practice, this may induce undesirable cross-compensation effects, whereby a discrepancy that should be attributed to one flavor is partially absorbed by shifting another, or the optimization becomes dominated by the statistically most abundant flavor in the sample. For this reason, it is more natural to define the flavor-conditional target distribution that the OT map actually needs to learn through a separate statistical procedure.

In this paper, we propose a procedure to extract the jet-flavor responses of the tagger directly from control-region data, using the expectation-maximization (EM) algorithm as its statistical foundation~\cite{em}. The EM algorithm is a standard iterative method for solving maximum likelihood problems with unobserved latent variables. In each iteration, it calculates the event-wise flavor probabilities under the current model and uses them to re-estimate the model parameters. In our problem, the flavor label corresponds to the latent variable, and data from multiple control regions are used simultaneously to jointly estimate the flavor component densities and the regional mixing fractions.
Since the NLL decreases monotonically during EM iterations~\cite{wu_em_convergence}, the procedure converges to a stationary point in problems where the NLL is bounded below. While a lower bound on the NLL is not guaranteed in full generality, in practice the use of a finite-capacity model to represent $q_f$ typically imposes an implicit upper bound on the pdf, and appropriate regularization prevents overfitting; it is therefore reasonable to treat the NLL as bounded below throughout the iterations.
The present study validates this procedure in a pseudo-data closure setup before application to collision data.

The core idea is to combine three control regions ($b$-enriched, $c$-enriched, and $\ell$-enriched) into a single mixture model, simultaneously estimating the density of each flavor component and the regional mixing fractions from the data. The specific control-region configurations and validation procedures are described in Section~\ref{sec:setup}. For an input variable $x$ that has undergone ILR transformation and common normalization, let the density of the flavor component $f\in\mathcal{F}=\{b,c,\ell\}$ be $q_f(x)$, and the mixing fraction in region $r\in \mathcal{R}=\{r_b,r_c,r_\ell\}$ be $\pi_{r,f}$. Then the observed distribution in each region is modeled as
\begin{equation}
p_r(x)=\sum_{f\in\mathcal{F}} \pi_{r,f}\, q_f(x)
\label{eq:monf_mixture_model}
\end{equation}
Throughout this paper, the following notation is used for flavor-conditional densities: $q_f^\star(x)$ denotes the truth-level (ideal) density, $\hat{q}_f(x)$ denotes the EM-fitted density at the converged solution, and $q_f^{\mathrm{MC}}(x)$ denotes the MC-supervised pretrained density used as initialization.

\paragraph{Transferability assumption.}
The following derivation relies on the assumption that the flavor-conditional
tagger response $q_f(x)$ is independent of the control region $r$,
within the phase-space bin under consideration:
\begin{equation}
    q_{r,f}(x) \equiv q_f(x).
\end{equation}
It states that, after fixing the relevant
kinematic and event-category variables, the control-region selection changes only
the flavor composition $\pi_{r,f}$, not the flavor-conditional shape of the tagger
response itself.

This assumption is not automatic. In particular, it can be violated if different
control regions populate different jet-$p_T$, $\eta$, or event-topology
distributions, since the tagger response may depend on these variables. Such
dependences must therefore be removed by performing the calibration in sufficiently
fine bins of the relevant phase-space variables, or by including those variables as
conditioning inputs to the normalizing-flow component densities.

In this work, we assume that this phase-space matching has been achieved. In the
simulation closure study presented here, the assumption holds by construction, since
both the MC and pseudo-data samples are drawn from the same jet population with
controlled kinematic distributions. Under
this assumption, the remaining difference among the control regions is described by
the mixture fractions $\pi_{r,f}$, while the component density $q_f(x)$ is common
across regions. This is the same transferability requirement underlying conventional
scale-factor calibrations: a calibration extracted in a control region can only be
applied elsewhere if the flavor-conditional response is stable after the chosen
binning or conditioning.

\paragraph{Practical identifiability.}
The decomposition in Eq.~\eqref{eq:monf_mixture_model} is stabilized by the following practical conditions:
(i)~the composition matrix $\Pi$ must be full-rank and well-conditioned, requiring that the control regions have sufficiently distinct flavor compositions;
(ii)~the component densities $q_f$ must be sufficiently distinguishable so that the flavor assignments are not degenerate;
(iii)~MC-supervised pretraining anchors the flavor-label semantics and prevents label switching.
This is therefore not treated as an unconstrained blind source separation problem; the flavor definitions are fixed by the pretraining stage.

Now, let the $i$-th data point in region $r$ be $x_{r,i}$. Since the flavor of this event is not directly observed, we introduce a latent variable $z_{r,i}\in\mathcal{F}$ to represent it. Then the joint probability of $x_{r,i}$ and $z_{r,i}$ is
\begin{equation}
p(x_{r,i},z_{r,i}=k)=\pi_{r,k}\,q_k(x_{r,i}),
\qquad k\in\mathcal{F}
\label{eq:joint_prob_em}
\end{equation}
Therefore, the complete-data likelihood for the observed values and the latent flavor labels is
\begin{equation}
\mathcal{L}
=
\prod_{r\in\mathcal{R}}\prod_i\prod_{k\in\mathcal{F}}
\left[\pi_{r,k}q_k(x_{r,i})\right]^{\delta_{z_{r,i},k}},
\label{eq:complete_data_likelihood}
\end{equation}
Taking the logarithm yields
\begin{equation}
\log\mathcal{L}
=
\sum_{r\in\mathcal{R}}\sum_i\sum_{k\in\mathcal{F}}
\delta_{z_{r,i},k}\left[\log\pi_{r,k}+\log q_k(x_{r,i})\right].
\label{eq:complete_data_loglik}
\end{equation}
The EM algorithm iteratively maximizes this expression averaged over the distribution of $z_{r,i}$ under the current model.

\paragraph{E-step.}
Since the latent variable $z_{r,i}$ is not directly observed, we define the probability that event $x_{r,i}$ belongs to flavor $k$ at the current iteration $t$ as
\begin{equation}
\begin{split}
\gamma_{r,i,k}^{(t)}
&:=
\mathbb{P}\!\left(z_{r,i}=k\mid x_{r,i}\right) \\
&=
\frac{\pi_{r,k}^{(t)}q_k^{(t)}(x_{r,i})}{\sum_{j\in\mathcal{F}}\pi_{r,j}^{(t)}q_j^{(t)}(x_{r,i})}
\end{split}
\label{eq:em_responsibility_local}
\end{equation}
Then the expected value of the log-likelihood with respect to these probabilities is
\begin{multline}
\mathbf{E}_{z\mid x;\,\pi^{(t)},q^{(t)}}\!\left[\log\mathcal{L}\right]
= \\
\sum_{r\in\mathcal{R}}\sum_i\sum_{k\in\mathcal{F}}
\gamma_{r,i,k}^{(t)}
\left[\log\pi_{r,k}+\log q_k(x_{r,i})\right]
\label{eq:em_q_function}
\end{multline}

\paragraph{M-step.}
In the next step, we maximize a regularized version of Eq.~\eqref{eq:em_q_function} with respect to the mixing fractions and the component densities. Rather than using the unregularized closed-form update for $\pi$, each row of the composition matrix is parameterized by free logits
\begin{equation}
    a_{r,j}=\log\frac{\pi_{r,j}}{\pi_{r,K}},
    \qquad j=1,\ldots,K-1,
\end{equation}
with the inverse map
\begin{align}
    D_r(a)&=1+\sum_{m=1}^{K-1}\exp(a_{r,m}), \nonumber\\
    \pi_{r,j}(a)&=\frac{\exp(a_{r,j})}{D_r(a)},
    \qquad j=1,\ldots,K-1, \nonumber\\
    \pi_{r,K}(a)&=\frac{1}{D_r(a)}.
\end{align}
We then introduce a Gaussian prior in this logit space,
\begin{equation}
    \log p(a)
    =
    -\frac{1}{2}
    \sum_{r\in\mathcal{R}}
    (a_r-a_r^{0})^{T}
    \Sigma_{a,r}^{-1}
    (a_r-a_r^{0})
    +\mathrm{const.},
    \label{eq:logit_prior}
\end{equation}
where $a_r^0$ is obtained from the nominal control-region composition. The resulting M-step is a maximum-a-posteriori update:
\begin{equation}
    (a^{(t+1)},\theta^{(t+1)})
    =
    \underset{a,\theta}{\operatorname{arg\,max}}
    \,\mathcal{J}^{(t)}(a,\theta),
    \label{eq:regularized_m_step}
\end{equation}
where
\begin{multline}
    \mathcal{J}^{(t)}(a,\theta)
    =
    \sum_{r,i,k}
    \gamma_{r,i,k}^{(t)}
    \Bigl(
    \log \pi_{r,k}(a_r)
    +\log q_k(x_{r,i};\theta_k)
    \Bigr) \\
    -\frac{1}{2}
    \sum_{r}
    (a_r-a_r^{0})^{T}
    \Sigma_{a,r}^{-1}
    (a_r-a_r^{0}).
\end{multline}
The logit prior is not meant to fix the mixture fractions to their nominal values; its width controls how far the data are allowed to move the regional compositions. We introduce this restriction because the mixture-normalizing-flow EM objective is highly non-convex and the negative log-likelihood can contain many local minima. Without a composition constraint, the fit can trade off $\Pi$ against the component densities $q_f$, leading to label switching or cross-compensation between flavors. The prior keeps the optimization in a physically plausible neighborhood while still allowing the data to update the mixture fractions.

This is a combined-sample objective: all control-region events enter the same normalizing-flow update, with their event-wise responsibilities providing the flavor weights, while the logit prior regularizes only the regional mixture fractions. The update of $q_f$ intentionally pools information from all regions, whereas the update of $a_r$ remains row-local apart from the shared dependence through the component-density parameters. In the closure study below, the pseudo-data regions are sampled with the same event budget and unit event weights; more general applications may introduce explicit per-region or per-event weights in the same objective.

\section{Conditioning and Composition-Induced Shape Response}
\label{sec:error}
While the EM algorithm monotonically increases the likelihood, the uniqueness of the solution is not guaranteed. In other words, the fitted mixing matrix $\Pi$ and component densities $q_f$ can converge to a stationary point of the likelihood rather than to the global maximum. For the present method, the most important practical issue is how uncertainty in the fitted mixture model propagates into the EM-derived target used by the OT map.

The leading stability requirement is that the regional composition matrix be well conditioned. Let
\begin{equation}
    P(x)=\Pi Q(x)
\end{equation}
denote the regional mixture relation, where $P(x)$ collects the regional densities and $Q(x)$ collects the flavor-conditional component densities. Linearizing around the ideal solution gives, schematically,
\begin{equation}
    \Delta P(x) \simeq \Pi^\star \Delta Q(x)+\Delta\Pi\,Q^\star(x).
\end{equation}
When $\Pi^\star$ is invertible,
\begin{equation}
    \Delta Q(x) \simeq
    (\Pi^\star)^{-1}
    \left[\Delta P(x)-\Delta\Pi\,Q^\star(x)\right].
\end{equation}
Thus density-fitting errors and composition errors are amplified by the inverse of the composition matrix. A useful diagnostic is the condition number
\begin{equation}
    \kappa(\Pi^\star)=\|(\Pi^\star)^{-1}\|\,\|\Pi^\star\|.
\end{equation}
If the control regions are high-purity and have sufficiently distinct compositions, $\Pi^\star$ is close to diagonal and $\kappa(\Pi^\star)$ is close to unity, so the extraction does not significantly amplify the underlying errors. Conversely, if two control regions have similar flavor compositions, $\Pi^\star$ becomes ill conditioned even if the leading-flavor fractions are sizable, and the uncertainty on $Q$ can grow rapidly.

For completeness, the corresponding pointwise perturbative bound can be written explicitly. Let $\|\cdot\|$ denote a submultiplicative matrix norm and its induced vector norm. At points $x$ where $\|P^\star(x)\|$ and $\|Q^\star(x)\|$ are nonzero, write
\begin{equation}
\begin{gathered}
    P(x)=P^\star(x)+\Delta P(x), \qquad
    \Pi=\Pi^\star+\Delta\Pi,\\
    Q(x)=Q^\star(x)+\Delta Q(x).
\end{gathered}
\end{equation}
Keeping the nonlinear perturbation term gives
\begin{equation}
    \Delta P(x)=
    \Pi^\star\Delta Q(x)
    +\Delta\Pi\,Q^\star(x)
    +\Delta\Pi\,\Delta Q(x).
\end{equation}
If $\Pi^\star$ is invertible and
\begin{equation}
    \|(\Pi^\star)^{-1}\|\,\|\Delta\Pi\|<1,
\end{equation}
then
\begin{equation}
    \|\Delta Q(x)\|
    \le
    \frac{\|(\Pi^\star)^{-1}\|
    \left(\|\Delta P(x)\|+\|\Delta\Pi\|\,\|Q^\star(x)\|\right)}
    {1-\|(\Pi^\star)^{-1}\|\,\|\Delta\Pi\|}.
    \label{eq:pointwise_q_bound}
\end{equation}
Equivalently, using $\|P^\star(x)\|\le \|\Pi^\star\|\,\|Q^\star(x)\|$, one obtains the diagnostic relative form
\begin{multline}
    \frac{\|\Delta Q(x)\|}{\|Q^\star(x)\|}
    \le
    \frac{\kappa(\Pi^\star)}
    {1-\kappa(\Pi^\star)\dfrac{\|\Delta\Pi\|}{\|\Pi^\star\|}}\\
    \times
    \left(
    \frac{\|\Delta P(x)\|}{\|P^\star(x)\|}
    +
    \frac{\|\Delta\Pi\|}{\|\Pi^\star\|}
    \right).
    \label{eq:relative_q_bound}
\end{multline}
This expression is used only as a local diagnostic: it is not intended to control regions where the true regional density vector is near zero, for which relative pointwise errors are inherently unstable.

For the practical uncertainty propagation, we use the enriched-region $\Delta\Pi$ proxy for the induced component-density perturbation. For each flavor $f$, let $r_f$ denote the corresponding flavor-enriched control region, Assuming the high-purity limit in each region, the component-density response is approximated by
\begin{equation}
    \Delta\log q_f(x)
    \simeq
    \sum_{h\neq f}
    \Delta\pi_{r_f,h}
    \left(1-\frac{q_h(x)}{q_f(x)}\right).
    \label{eq:delta_pi_proxy}
\end{equation}
Equivalently, multiplying by $q_f(x)$ gives the density-level proxy
\begin{equation}
    \Delta q_f(x)
    \simeq
    \sum_{h\neq f}
    \Delta\pi_{r_f,h}
    \left(q_f(x)-q_h(x)\right).
    \label{eq:delta_pi_density_proxy}
\end{equation}
Equation~\eqref{eq:delta_pi_proxy} gives the practical interpretation of the proxy. First, the relevant quantity is the absolute error on the background fractions in the $f$-enriched region, not the relative uncertainty of a small background component. A high-purity region can therefore remain stable even when the fractional uncertainty on a tiny contamination is large, provided the absolute contamination error is small. Second, the induced shape error decreases as the tagger becomes more discriminating. In regions of score space where flavor $f$ dominates and the competing densities $q_h(x)$ are small, the ratio $q_h(x)/q_f(x)$ is suppressed and the perturbation mainly changes the normalization of the extracted component. The most dangerous regions are instead those where $q_h(x)/q_f(x)$ is large, corresponding to ambiguous or misidentified score regions. There, an error in the off-diagonal composition can directly inject the shape of the wrong flavor into the inferred $q_f$. The operator form of this proxy, evaluated at the fitted densities, is introduced in Section~\ref{sec:syst} as Eq.~\eqref{eq:A_proxy_definition}.

\section{Systematic Uncertainty Prescription}
\label{sec:syst}

A rigorous evaluation of the systematic uncertainties intrinsic to this method would
in principle require the full covariance matrix over both the mixing parameters
$\Pi$ and the normalizing-flow parameters. However, since the model parameters
typically number in the tens to hundreds of thousands, computing and inverting the
full covariance matrix over all variables is not feasible due to numerical stability
and computational constraints. A direct bootstrap treatment would also require many
repeated EM and OT training runs, which is computationally prohibitive. We therefore adopt a pragmatic uncertainty treatment that focuses on the dominant uncertainty induced by the fitted mixture composition.

The basic idea is to use the composition-induced response in
Eq.~\eqref{eq:delta_pi_proxy} as a local proxy for the uncertainty band of the
extracted component densities $q_f$. The remaining practical question is what
magnitude of perturbation $\Delta\pi$ should be assigned. If $\Delta\pi$ is chosen
too small, the resulting band will underestimate the effect; if it is chosen too large,
the band becomes unnecessarily conservative. In the present work, we take as a starting
point the local curvature of the mixture likelihood at the converged solution under the
fixed-$q_f$ approximation, and then use a linearized feedback operator as a diagnostic
model for how such composition perturbations may be amplified through the EM cycle.

In this section, fitted quantities are denoted by hats. In particular,
$\hat q_f$, $\hat\pi_{r,f}$, and $\hat a_r$ denote the component density,
mixture fraction, and mixture-logit parameters obtained at the converged EM solution.
The fitted regional density and the fitted responsibility are
\begin{align}
    \hat p_r(x)
    &=
    \sum_{k\in\mathcal{F}}
    \hat\pi_{r,k}\,\hat q_k(x),
    \label{eq:fitted_regional_density}\\[3pt]
    \hat\gamma_{r,i,k}
    &=
    \frac{
        \hat\pi_{r,k}\,\hat q_k(x_{r,i})
    }{
        \sum_{j\in\mathcal{F}}
        \hat\pi_{r,j}\,\hat q_j(x_{r,i})
    }.
    \label{eq:fitted_responsibility}
\end{align}
The superscript $\star$, used elsewhere in this paper, is reserved for ideal or
truth-level quantities and is not used for the fitted point in this section.

In the fixed-$\hat q_f$ approximation, the normalizing-flow component densities are
held fixed at their fitted values and only the mixture logits $a$ are varied.
Without the logit prior, the relevant observed-data negative log-likelihood is
\begin{equation}
    \mathcal{L}_{\mathrm{mix}}(a;\hat q)
    =
    -\sum_{r}\sum_i
    \log\left[
        \sum_{k\in\mathcal{F}}
        \pi_{r,k}(a_r)\,
        \hat q_k(x_{r,i})
    \right].
    \label{eq:mixing_nll_fixed_q}
\end{equation}
The local covariance of the free mixture-logit parameters is then approximated by
the inverse Hessian evaluated at the fitted point:
\begin{equation}
    V_a^{(0)}
    =
    \left[
    \left.
    \frac{\partial^2 \mathcal{L}_{\mathrm{mix}}(a;\hat q)}
    {\partial a\,\partial a^{T}}
    \right|_{a=\hat a}
    \right]^{-1}.
    \label{eq:initial_logit_covariance_no_prior}
\end{equation}
This covariance should be interpreted as the initial composition uncertainty before
accounting for the feedback between the mixture fractions and the extracted
component densities. If a logit prior is included, the same expression can be
generalized by adding the prior precision to the Hessian,
\begin{equation}
    V_a^{(0)}
    =
    \left[
    \left.
    \frac{\partial^2 \mathcal{L}_{\mathrm{mix}}(a;\hat q)}
    {\partial a\,\partial a^{T}}
    \right|_{a=\hat a}
    + \Sigma_a^{-1}
    \right]^{-1}.
    \label{eq:initial_logit_covariance_with_prior}
\end{equation}
In the present diagnostic, however, we first quote the no-prior covariance in
Eq.~\eqref{eq:initial_logit_covariance_no_prior} in order to isolate how strongly the
data themselves constrain the mixture composition.

The perturbation of the mixture fractions induces a perturbation of the component
densities, which in turn modifies the E-step responsibilities and hence the next
mixture-fraction update. We approximate this closed-loop response by a linear
operator $F$ around the fitted solution $(\hat\pi,\hat q)$. The purpose of
$F$ is not to provide an exact model of normalizing-flow retraining, but rather to
diagnose how composition fluctuations are amplified or damped by the EM feedback.

\paragraph{Closed-form M-step without prior.}
To construct a simple local feedback model, we first consider the idealized case in
which the logit prior is absent. In that limit, the M-step for the mixing fractions
reduces to a closed-form update. For each region $r$, we maximize
\begin{equation}
    \sum_i\sum_{k\in\mathcal{F}} \gamma_{r,i,k}^{(t)}\log \pi_{r,k}
    + \lambda_r\!\left(1-\sum_{k\in\mathcal{F}}\pi_{r,k}\right)
\end{equation}
subject to $\sum_k \pi_{r,k}=1$. Taking the derivative with respect to
$\pi_{r,k}$ gives
\begin{equation}
    \sum_i\frac{\gamma^{(t)}_{r,i,k}}{\pi_{r,k}}-\lambda_r=0.
\end{equation}
Multiplying by $\pi_{r,k}$ and summing over $k$, then using the normalization
constraint, yields
\begin{equation}
    \lambda_r = \sum_k\sum_i\gamma^{(t)}_{r,i,k}.
\end{equation}
Therefore, the closed-form update rule is
\begin{equation}
    \pi_{r,k}^{(t+1)}
    =
    \frac{\sum_i\gamma^{(t)}_{r,i,k}}
    {\sum_j\sum_i \gamma^{(t)}_{r,i,j}}
    =
    \frac{1}{N_r}\sum_i\gamma^{(t)}_{r,i,k},
    \label{eq:pi_closed_form}
\end{equation}
where $N_r$ is the number of events in region $r$. Thus, the updated mixture
fraction equals the average responsibility of flavor $k$ in that region.

\paragraph{Derivation of the feedback operator.}
We use Eq.~\eqref{eq:pi_closed_form} to derive the linearized one-step feedback
around the fitted solution. Writing
\begin{equation}
    \pi_{r,k}=\hat\pi_{r,k}+\delta\pi_{r,k},
\end{equation}
the perturbed E-step responsibility is
\begin{equation}
    \delta\gamma_{r,i,k}
    =
    \underbrace{
      \sum_m
      \frac{\hat q_m(x_{r,i})}{\hat p_r(x_{r,i})}
      \bigl(\delta_{km}-\hat\gamma_{r,i,k}\bigr)
      \delta\pi_{r,m}
    }_{\text{direct response}}
    +
    \underbrace{
      \sum_j
      \hat\gamma_{r,i,k}
      \bigl(\delta_{kj}-\hat\gamma_{r,i,j}\bigr)
      \delta\!\log q_j(x_{r,i})
    }_{\text{density-induced response}} .
    \label{eq:delta_gamma_linearized}
\end{equation}
Applying the closed-form M-step~\eqref{eq:pi_closed_form} at linear order gives
\begin{equation}
    \delta\pi_{r,k}^{\mathrm{new}}
    =
    \frac{1}{N_r}\sum_i \delta\gamma_{r,i,k}.
    \label{eq:delta_pi_new}
\end{equation}
The direct-response matrix $B$ and the density-coupling matrix $C$ are therefore
\begin{align}
    B_{r,k,m}
    &=
    \frac{1}{N_r}\sum_i
    \frac{\hat q_m(x_{r,i})}{\hat p_r(x_{r,i})}
    \bigl(\delta_{km}-\hat\gamma_{r,i,k}\bigr),
    \label{eq:B_matrix}\\[4pt]
    C_{r,k,j}
    &=
    \frac{1}{N_r}\sum_i
    \hat\gamma_{r,i,k}
    \bigl(\delta_{kj}-\hat\gamma_{r,i,j}\bigr).
    \label{eq:C_matrix}
\end{align}

To close this local response model, we now introduce a proxy operator that maps a
composition perturbation into a perturbation of the component log-density. For an event $x_{r,i}$, the proxy
response is
\begin{equation}
    \left[A_{\mathrm{proxy}}\delta\pi\right]_j(x_{r,i})
    \equiv
    \delta\!\log q_j(x_{r,i})
    =
    \sum_{h\neq j}
    \delta\pi_{r_j,h}
    \left(
        1-\frac{\hat q_h(x_{r,i})}{\hat q_j(x_{r,i})}
    \right),
    \label{eq:A_proxy_definition}
\end{equation}
where $r_j$ denotes the control region enriched in flavor $j$. This is the
operator form of Eq.~\eqref{eq:delta_pi_proxy}, evaluated at the fitted component
densities. Substituting Eq.~\eqref{eq:A_proxy_definition} into
Eq.~\eqref{eq:delta_gamma_linearized} gives
\begin{equation}
    \delta\pi^{\mathrm{new}}
    =
    \left(B+C\,A_{\mathrm{proxy}}\right)\delta\pi .
    \label{eq:pi_feedback_operator}
\end{equation}

Finally, we express the response in the free-logit parameterization.  Let $J$
denote the Jacobian that maps an infinitesimal variation of the free logits to the
corresponding variation of the mixture fractions on the simplex.  Evaluated at the
fitted mixture fractions, it is given by
\begin{equation}
    J_{r,k,j}
    =
    \left.
    \frac{\partial\pi_{r,k}}{\partial a_{r,j}}
    \right|_{a=\hat a}
    =
    \hat\pi_{r,k}
    \left(\delta_{kj}-\hat\pi_{r,j}\right),
    \qquad j=1,\ldots,K-1 .
    \label{eq:softmax_jacobian}
\end{equation}
Thus, for a small logit perturbation $\delta a_r$, the induced mixture-fraction
perturbation is
\begin{equation}
    \delta\pi_r = J_r\,\delta a_r .
\end{equation}

Conversely, a mixture-fraction perturbation satisfying
$\sum_k \delta\pi_{r,k}=0$ can be expressed in terms of the corresponding
free-logit perturbation by differentiating
$a_{r,j}=\log(\pi_{r,j}/\pi_{r,K})$.  This gives
\begin{equation}
    [\delta a_r]_j
    =
    \frac{\delta\pi_{r,j}}{\hat\pi_{r,j}}
    -
    \frac{\delta\pi_{r,K}}{\hat\pi_{r,K}},
    \qquad j=1,\ldots,K-1 .
    \label{eq:M_operator}
\end{equation}
We denote this linear conversion from $\delta\pi_r$ to $\delta a_r$ by $M_r$.
With this notation, the one-step response in the free-logit coordinates is
\begin{equation}
    \delta a^{\mathrm{new}}
    =
    M\left(B+C\,A_{\mathrm{proxy}}\right)J\,\delta a
    \equiv
    F\,\delta a .
    \label{eq:F_operator}
\end{equation}

If the feedback is contractive, repeated composition-density feedback gives the
Neumann series
\begin{equation}
    \delta a_{\mathrm{eff}}
    =
    \left(I+F+F^2+\cdots\right)\delta a
    =
    (I-F)^{-1}\delta a .
    \label{eq:neumann_feedback}
\end{equation}
Since $F$ is not guaranteed to be symmetric, we do not interpret its
eigenvectors as orthogonal uncertainty modes. Instead, we use the spectral radius
$\rho(F)$ as a convergence diagnostic and the singular values of
$(I-F)^{-1}$ as a robust measure of perturbation amplification. In particular, if
$\rho(F)<1$, the linearized feedback is locally contractive and the corresponding effective covariance model is 
\begin{equation}
    V_a^{\mathrm{eff}}
    =
    (I-F)^{-1}
    V_a^{(0)}
    (I-F)^{-T}.
    \label{eq:effective_logit_covariance}
\end{equation}

Within this proxy model, the corresponding component-density uncertainty is represented by propagating this
covariance through the softmax Jacobian and the proxy map:
\begin{equation}
    \mathrm{Cov}\!\left[\delta\!\log q(x)\right]
    \simeq
    A_{\mathrm{proxy}}(x)\,
    J\,V_a^{\mathrm{eff}}\,J^{T}
    A_{\mathrm{proxy}}(x)^{T}.
    \label{eq:q_covariance_proxy}
\end{equation}
Equivalently, one may draw variations $\delta a$ from
$V_a^{\mathrm{eff}}$, convert them to $\delta\pi=J\delta a$, and evaluate
Eq.~\eqref{eq:A_proxy_definition} to obtain up/down variations of the extracted
component densities.

If $\rho(F)\ge 1$, the Neumann expansion does not converge. More generally,
even when $\rho(F)<1$, individual singular channels of
$G=(I-F)^{-1}$ may be excessively amplified. Such directions indicate that the
composition-density decomposition is not sufficiently stabilized by this problem setup alone. 
We therefore do not assign their uncertainty using the formal feedback
factor $G$ alone. Instead, these directions are treated as prior- or
regularization-dominated and are covered by an externally specified composition
variation.

Since $F$ is not guaranteed to be normal, the input and output response
directions are not generally the same. Writing
\begin{equation}
    G = U S V^{T},
\end{equation}
a perturbation along the right singular vector $v_i$ is mapped to an output
response along the left singular vector $u_i$,
\begin{equation}
    G v_i = \sigma_i u_i .
\end{equation}

In the numerical studies presented in Sec.~\ref{sec:results}, the dominant
amplified channels are found to have nearly aligned left and right singular
vectors, $ |u_i^T v_i| \simeq 1 $. This indicates that, for the fitted
examples considered here, the dangerous input perturbation directions and the
corresponding output response directions are approximately the same. Nevertheless,
the prescription is formulated in terms of the left singular vectors, because
$V_a^{\mathrm{eff}}$ is an output-space covariance and this definition remains
well defined for a non-normal feedback operator.
Accordingly, the projected covariance prescription below is formulated in the
output logit space, using the left singular vectors. Let
\begin{equation}
    P_{\mathrm{st}} = U_{\mathrm{st}}U_{\mathrm{st}}^{T},
    \qquad
    P_{\mathrm{pr}} = U_{\mathrm{pr}}U_{\mathrm{pr}}^{T},
\end{equation}
where $U_{\mathrm{st}}$ contains the left singular vectors associated with
stable channels and $U_{\mathrm{pr}}$ contains those associated with
non-contractive or excessively amplified channels. A practical projected covariance prescription is then
\begin{equation}
    V_a^{\mathrm{eff}}
    =
    P_{\mathrm{st}}\,
    G V_a^{(0)} G^{T}\,
    P_{\mathrm{st}}
    +
    P_{\mathrm{pr}}\,
    V_a^{\mathrm{prior}}\,
    P_{\mathrm{pr}} .
    \label{eq:effective_logit_covariance_projected}
\end{equation}
Here $V_a^{\mathrm{prior}}$ is the logit-space covariance corresponding to the
external composition prior. In this work it is taken to be diagonal in the
free-logit coordinates, with widths chosen to reproduce the nominal composition
variation. The prior-dominated term in
Eq.~\eqref{eq:effective_logit_covariance_projected} should be interpreted as
conditional on this external variation being sufficiently conservative to cover
plausible deviations of the true mixture composition in the unstable directions.

In the implementation, Eq.~\eqref{eq:A_proxy_definition} is evaluated with the fitted
component densities at the converged EM solution. It is therefore used as a local
diagnostic and uncertainty-propagation proxy rather than as an exact model of
normalizing-flow retraining. The procedure captures the leading effect by which
mixing-fraction fluctuations can inject the shape of one flavor component into
another through the EM responsibility update.

\section{Simulation Samples and Tagger}
\label{sec:setup}
To validate the above method, we constructed a simulation-based closure study. The base sample is the $t\bar{t}$ heavy-flavour tagging dataset published on Zenodo~\cite{pond_dataset}; details of the sample composition and jet kinematics are given therein. The GN2 tagger~\cite{gn2} was trained on this sample using the Salt framework~\cite{salt}. To generate pseudo-data, we apply Gaussian smearing and random dropout to a subset of the input track features as a controlled toy mismodelling. The original sample is referred to as MC and the degraded sample as Pseudo-Data throughout.

The flavor-conditional densities $q_f$ in the EM procedure are each modeled by a neural spline flow using rational quadratic spline (RQS) transformations~\cite{rqs}. The OT maps $T_f$ are parameterized as gradients of input convex neural networks (ICNNs)~\cite{icnn,makkuva}, following the Brenier-OT framework.

\section{Analysis Procedure}
The analysis procedure is as follows. First, we construct three pseudo-data control regions using flavor-labeled samples. For each region, the mixture fraction is artificially set to be enriched in a specific flavor; the configuration used in this text is shown in Table~\ref{tab:pseudodata_regions}.

\begin{table*}[t]
\centering
\begin{tabular}{lccc}
\hline
Region & $f_b$ & $f_c$ & $f_\ell$ \\
\hline
$\ell$-enriched & 5.05\% & 8.87\% & 86.1\% \\
$c$-enriched & 5\% & 88.86\% & 6.14\% \\
$b$-enriched & 90.0\% & 4.0\% & 6.0\% \\
\hline
\end{tabular}
\caption{Representative flavor fractions used to construct the artificial
pseudo-data mixtures in the closure study.}
\label{tab:pseudodata_regions}
\end{table*}

For each pseudo-data sample, the flavor mixture is artificially fixed to a
representative enriched composition, as shown in Table~\ref{tab:pseudodata_regions}.
These values are chosen to be broadly inspired by the flavor-purity scales reported
in the CMS charm-jet calibration~\cite{cms_cjet_calib}, but are not intended to
reproduce the corresponding control regions event by event.

For the $c$-enriched sample, the opposite-sign minus same-sign dimuon subtraction used in Ref.~\cite{cms_cjet_calib} is not straightforward to emulate on an event-by-event basis in our pseudo-data construction. We therefore adopt a slightly reduced effective $c$-purity. For the $b$-enriched sample, although Ref.~\cite{cms_cjet_calib} uses both semileptonic and dileptonic $t\bar t$
categories, a dileptonic-like selection is expected to provide a cleaner $b$-enriched sample. We therefore assume a somewhat higher effective b-purity.

Truth-flavor labels are used only when constructing the pseudo-data control regions (to control mixture fractions) and during MC-supervised component pretraining. The EM extraction stage itself treats flavor labels as latent variables and operates only on unlabeled pseudo-data.

Subsequently, the data in each region undergo ILR-related preprocessing to be converted into EM inputs. We use MC truth labels only to pretrain each flavor component flow, and these pretrained flows are then used as the initialization for the EM cycle on unlabeled pseudo-data. Once the EM algorithm converges, we construct the empirical target measure $\hat{q}_f$ for each flavor, and finally train the OT map to transport the MC flavor source to this target. Thus, the overall workflow can be understood in five steps: (i)~pseudo-data region construction, (ii)~component pretraining, (iii)~data-based EM target extraction, (iv)~OT map training, and (v)~closure and transfer validation. For EM training, we initialize $\pi_{r,k}$ with random perturbations so that the setup reflects the realistic case in which the exact initial values are not known.
Specifically, for each region the initial composition is drawn by randomly selecting a perturbation in which the relative change of each component fraction does not exceed 20\%. 
The perturbation magnitude is further constrained such that, in the region enriched in flavor $f$, the prior width on the background-to-signal log-ratio is set to 0.3, i.e.\ $\Delta \log(f^\prime/f)=0.3$.
\section{Results}
\label{sec:results}

This section presents the toy-based validation results for the proposed calibration framework, proceeding through three stages: convergence and component extraction from the EM training, the trained OT transport maps, and finally closure on an independent validation region.

\subsection{EM Training and Component Extraction}

\begin{figure}[h!]
\centering
\includegraphics[width=0.48\columnwidth]{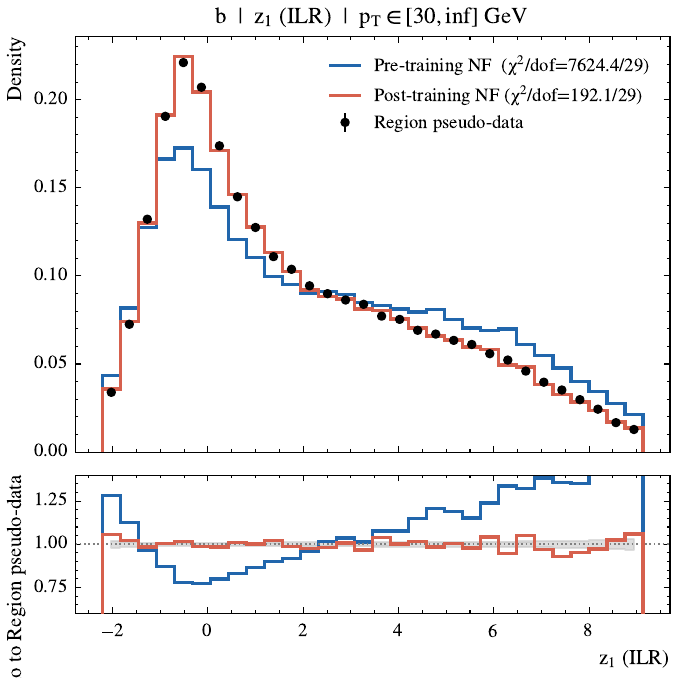}
\hfill
\includegraphics[width=0.48\columnwidth]{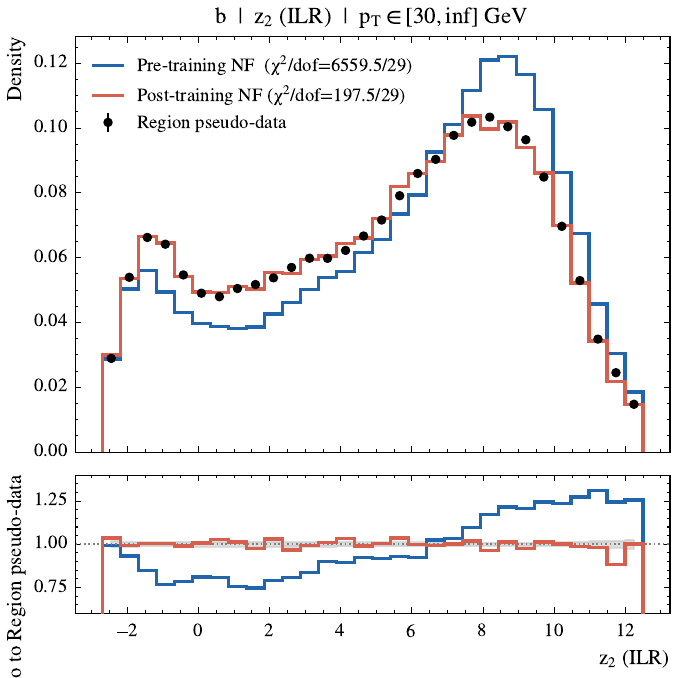}
\caption{Tagger output distributions before (pre-training) and after EM convergence in the $b$-enriched control region. The two panels correspond to the two tagger output features.}
\label{fig:EM_prepost}
\end{figure}

\paragraph{Convergence of the normalizing flows.}
Figure~\ref{fig:EM_prepost} shows the tagger output distributions before and after the EM cycle, using the $b$-enriched control region as a representative example. After convergence, the normalizing flow model for each flavor component provides a good description of the  pseudo-data in all three control regions, indicating that the EM cycle reaches a well-fitted solution in this pseudo-data study.

\paragraph{Fitted mixture fractions.}

\begin{figure}[h!]
\centering
\includegraphics[width=0.7\columnwidth]{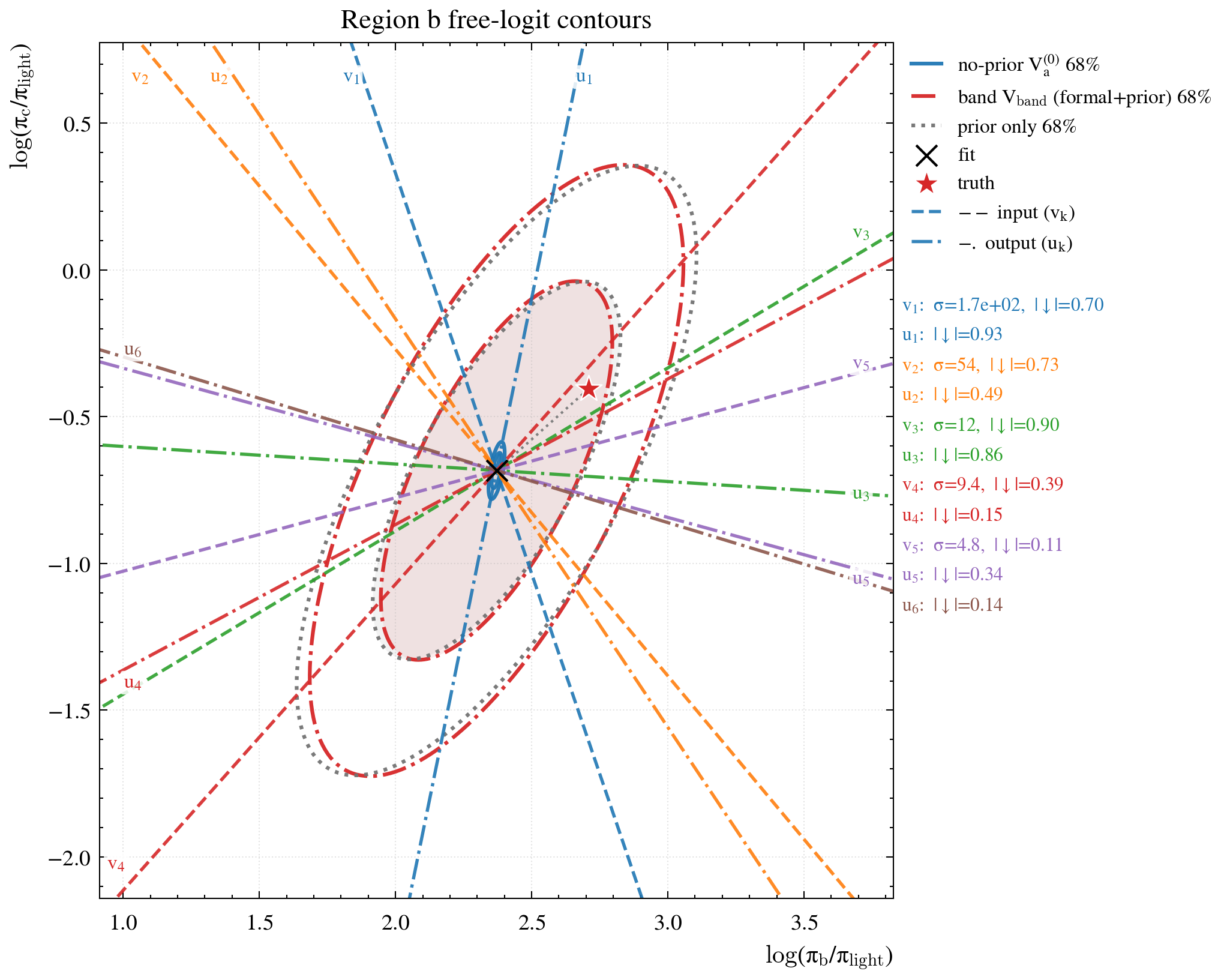}
\caption{Mixture-fraction logit plane and error contours for the $b$-enriched control region. Blue and red contours show the statistical and effective (prior-included) uncertainties, respectively. The fitted mixture fractions (crosses) and pseudo-data truth values (stars) are overlaid.}
\label{fig:logit_contour_b}
\end{figure}
\begin{figure}[h!]
\centering
\includegraphics[width=0.7\columnwidth]{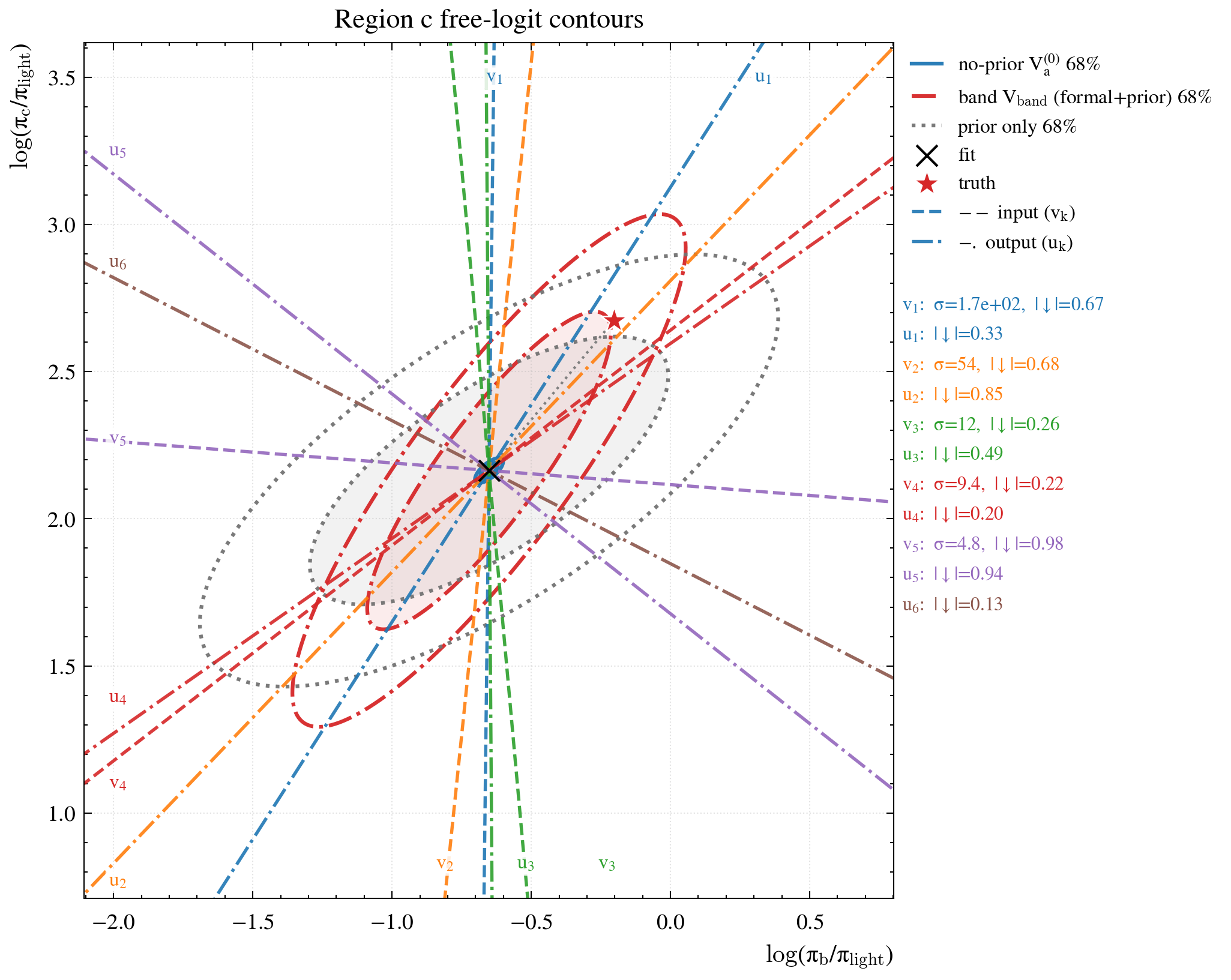}
\caption{Same as Fig.~\ref{fig:logit_contour_b} for the $c$-enriched control region.}
\label{fig:logit_contour_c}
\end{figure}
\begin{figure}[h!]
\centering
\includegraphics[width=0.7\columnwidth]{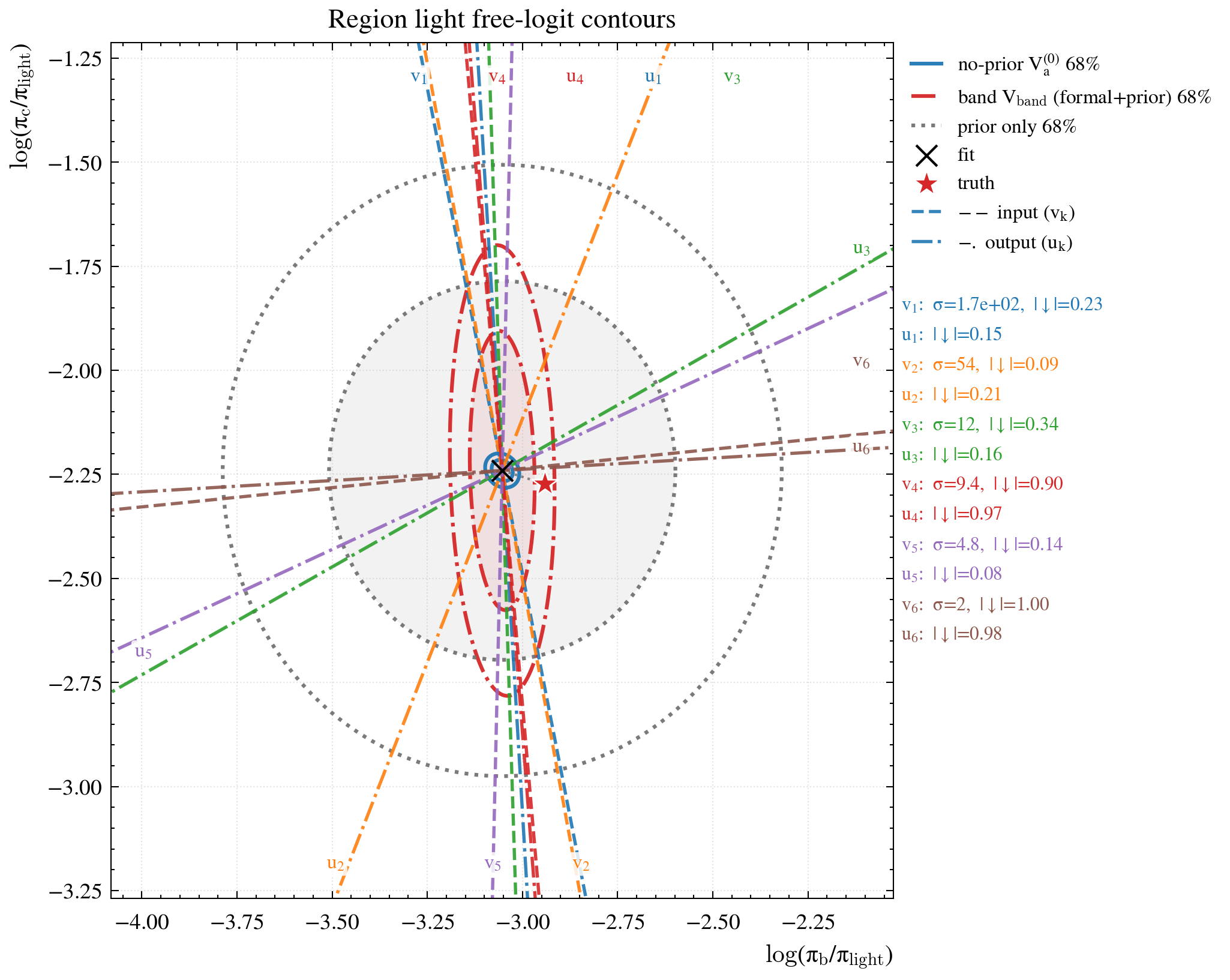}
\caption{Same as Fig.~\ref{fig:logit_contour_b} for the $\ell$-enriched control region.}
\label{fig:logit_contour_l}
\end{figure}
Figures~\ref{fig:logit_contour_b}--\ref{fig:logit_contour_l} display the mixture-fraction logit space and error contours for each of the three enriched control regions. The blue contours correspond to the statistical uncertainty on the mixture fractions derived from Eq.~(\ref{eq:initial_logit_covariance_no_prior}) given the current estimate of $q_f$, while the red contours show the effective uncertainty after incorporating the prior, as given by Eq.~(\ref{eq:effective_logit_covariance_projected}). The six input singular directions $v_1$--$v_6$, and output singular directions $u_1$--$u_6$ of the feedback operator $G = (1-F)^{-1}$ from Eq.~(\ref{eq:neumann_feedback}) span the full mixture-fraction space: with three regions each contributing three free parameters subject to one sum-to-unity constraint per region, the total degrees of freedom are six. The projections of each singular basis onto the corresponding regional subspace are labeled $|\!\downarrow\!|$; a value of unity indicates the direction lies entirely within that subspace. Entries smaller than 0.05 are suppressed.

In all three regions, the truth mixture fractions fall within approximately one sigma of the red contours. However, the singular value $\sigma_1$ associated with direction $v_1$ is substantially larger than the others, indicating that a perturbation along $v_1$ generates a large feedback response along $u_1$. This reflects a flat direction in the likelihood — an inherent degeneracy of jointly estimating $q_f$ and $\pi_{r,f}$ from data without additional constraints. Consequently, the proposed method should not be interpreted as simultaneously measuring both quantities from data. Rather, it provides a fitted estimate of $q_f$ whose local uncertainty is summarized by the effective covariance model. When the fitted mixture fractions drift significantly beyond the assumed prior, this should be understood as the EM training sliding along an unstable direction, not as a data-driven preference for a different composition. In practice, the prior must be set conservatively around the best available estimate of the true mixture fractions, and the fitted values should be verified to remain within acceptable deviations from the prior. Within this regime, the uncertainty prescription of Section~\ref{sec:syst} provides bands that remain bounded by the prior width in unconstrained directions while becoming narrower in directions where the data provide genuine discriminating power.

\paragraph{Extracted flavor-conditional distributions.}
\begin{figure}[h!]
\centering
\includegraphics[width=0.8\columnwidth]{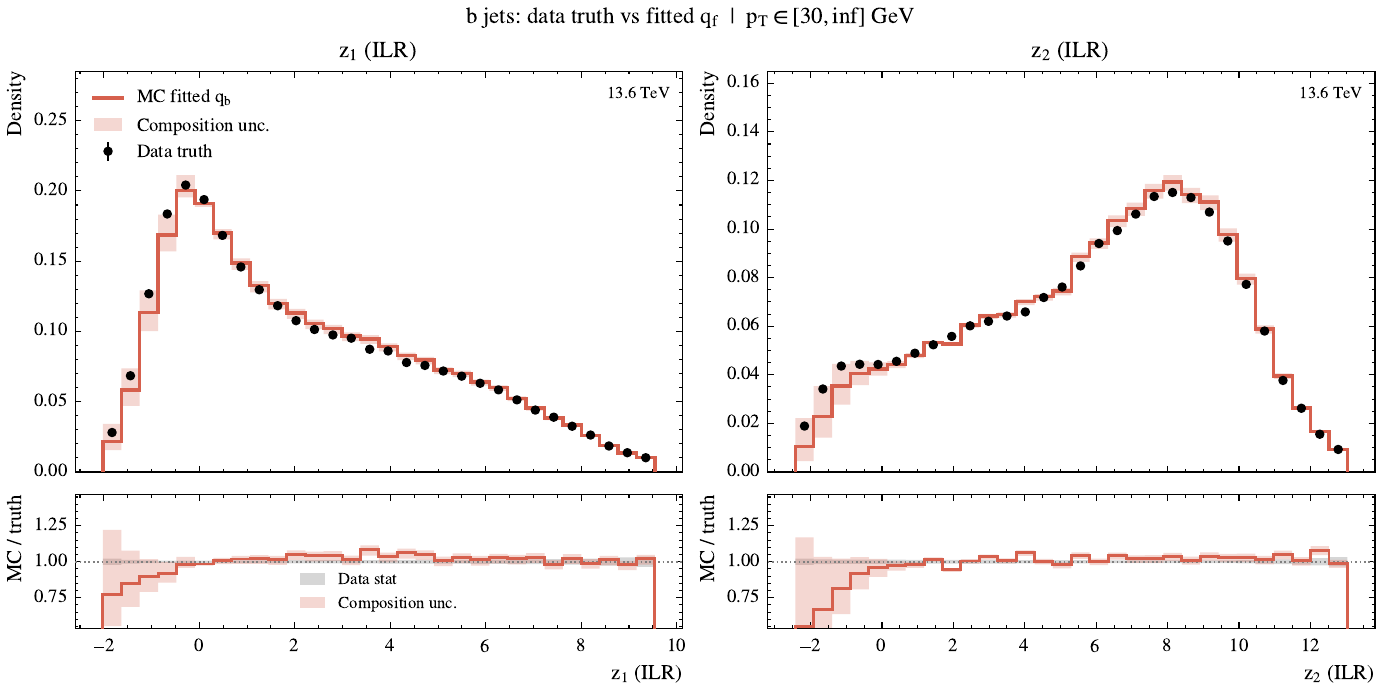}
\caption{Extracted $\hat{q}_f$ vs.\ truth $q^\star_f$ in the $b$-enriched control region, projected onto each tagger output feature. Error bands are derived from the effective covariance.}
\label{fig:qf_truth_vs_fitted_b}
\end{figure}

\begin{figure}[h!]
\centering
\includegraphics[width=0.8\columnwidth]{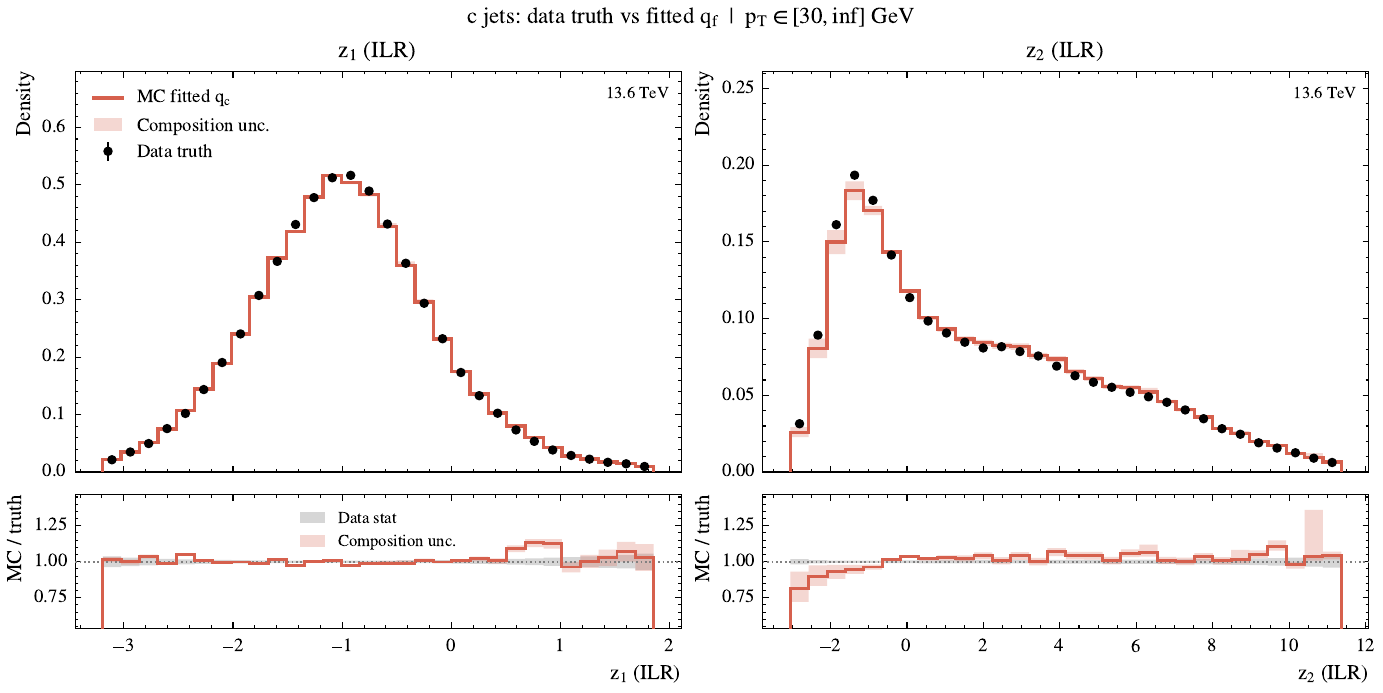}
\caption{Same as Fig.~\ref{fig:qf_truth_vs_fitted_b} for the $c$-enriched control region.}
\label{fig:qf_truth_vs_fitted_c}
\end{figure}

\begin{figure}[h!]
\centering
\includegraphics[width=0.8\columnwidth]{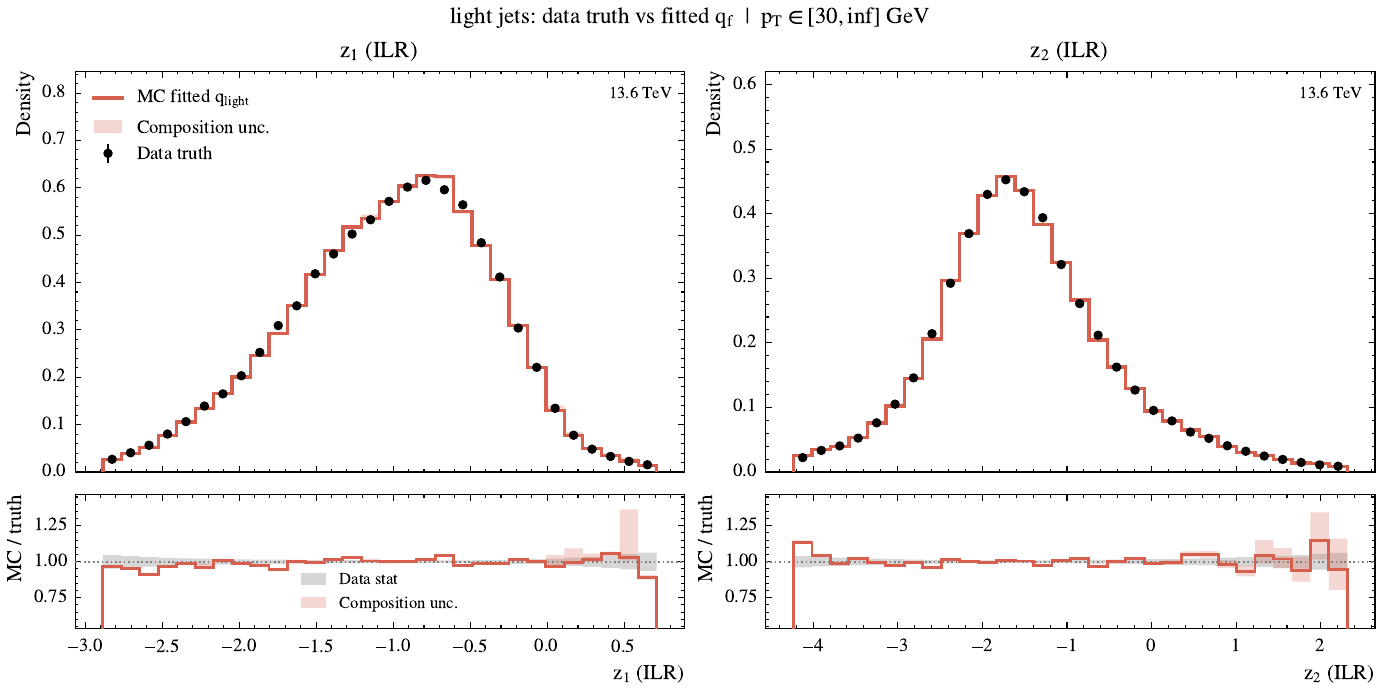}
\caption{Same as Fig.~\ref{fig:qf_truth_vs_fitted_b} for the $\ell$-enriched control region.}
\label{fig:qf_truth_vs_fitted_light}
\end{figure}

Figures~\ref{fig:qf_truth_vs_fitted_b}--\ref{fig:qf_truth_vs_fitted_light} compare the extracted $\hat{q}_f$ against the pseudo-data truth $q^\star_f$ for each control region, with each distribution projected onto the two tagger output features. In most regions the two agree well, and where deviations occur the associated error bands — estimated by propagating the effective covariance from Eq.~(\ref{eq:effective_logit_covariance_projected}) through the proxy $A$ of Eq.~(\ref{eq:A_proxy_definition}) are generally broad enough to cover the observed truth-level deviations. This outcome is consistent with the setup of the present study, in which the per-component composition uncertainty is at most 20\% and the prior is set conservatively at 30\%, ensuring that the prior-regularized bands remain broad enough to enclose the truth while the profiling step tightens them in constrained directions.

\subsection{Results of OT Map Training}

\begin{figure}[h!]
\centering
\includegraphics[width=0.9\columnwidth]{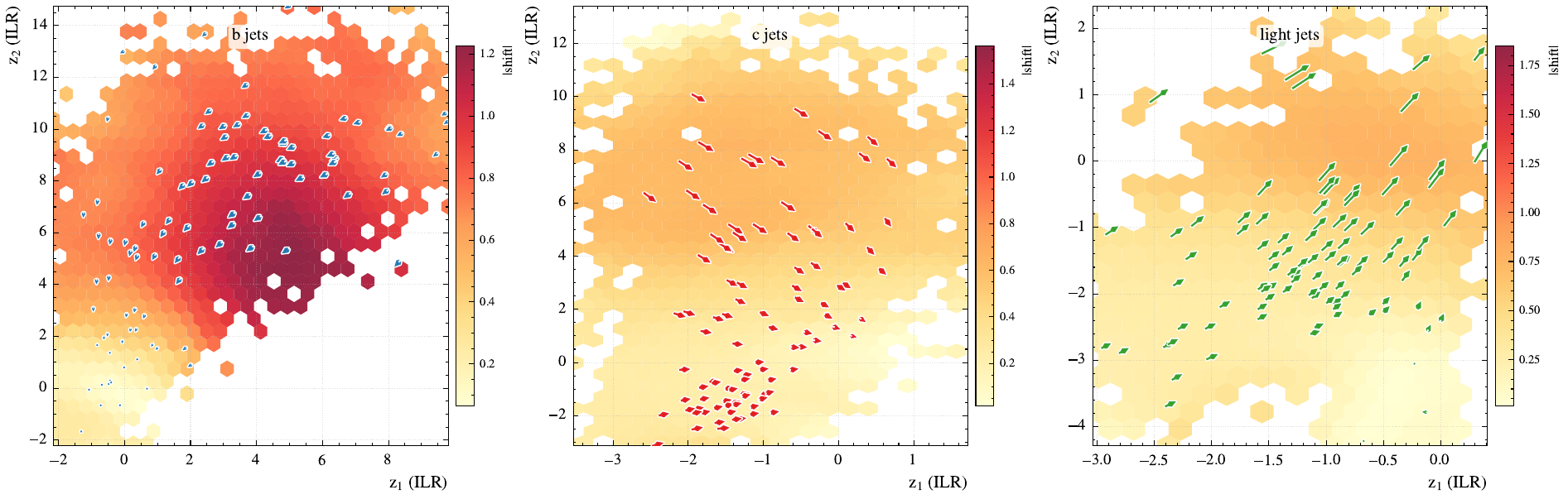}
\caption{Learned transport map visualized in the tagger output space. Arrows show the displacement of $b$-, $c$-, and light-flavor jets under the calibration map.}
\label{fig:ot_ilr}
\end{figure}

\begin{figure}[h!]
\centering
\includegraphics[width=0.9\columnwidth]{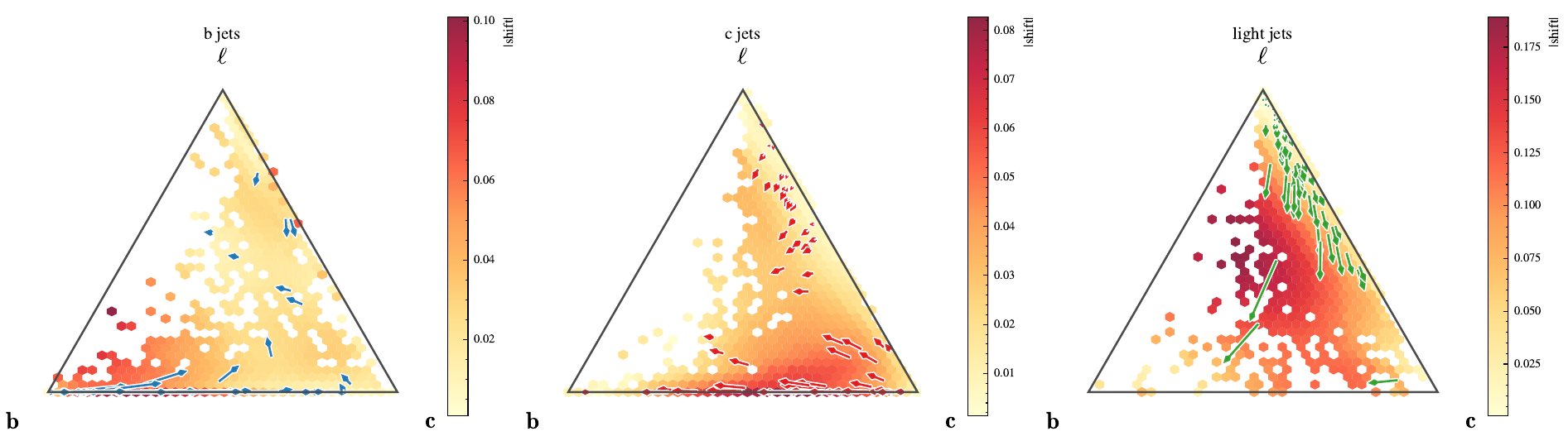}
\caption{Same transport map visualized on the probability simplex. Each vertex represents the tagger being fully confident in the corresponding flavor label.}
\label{fig:ot_simplex}
\end{figure}

Using the extracted $\hat{q}_f$ as targets, flavor-factorized OT maps are trained to transport the MC source distribution $q_f$ onto $\hat{q}_f$ for each flavor. Figure~\ref{fig:ot_ilr} shows the learned transport map in the two-dimensional tagger output space, with arrows indicating the displacement of jets with truth labels $b$, $c$, and $\ell$ under the map. The map is trained to minimize the squared Euclidean cost in this space, which corresponds to minimizing the Aitchison distance on the probability simplex. Figure~\ref{fig:ot_simplex} shows the same map visualized directly on the simplex, where each vertex corresponds to the tagger assigning 100\% probability to the corresponding flavor.

In both representations, the maps consistently shift jets away from the vertices, indicating that the calibrated tagger assigns lower confidence scores than the uncorrected MC. This is physically consistent with the pseudo-data tagger exhibiting lower discrimination power than the MC. In the language of traditional scale-factor measurements, an arrow pointing away from the $b$ vertex at a given score corresponds to a scale factor below unity at that operating point.
\subsection{Closure on Validation Region}

To assess whether the trained OT maps generalize beyond the enriched control regions used during EM training, we evaluate closure on an independent \emph{validation region} constructed as an equal mixture of $b$-, $c$-, and light-flavor jets ($\pi_{b}:\pi_{c}:\pi_{\ell} = 1:1:1$). This composition is deliberately agnostic to any flavor enrichment, and the region is not seen at any stage of the EM extraction or OT map training.

Closure is assessed on derived discriminant scores that are the primary quantities of interest in downstream analyses. Specifically, we examine the binary discriminants
\begin{equation}
    D_{b/c}(\kappa) \;=\; \frac{p_b}{p_b + \kappa\, p_c}, \qquad
    D_{c/\ell}(\kappa) \;=\; \frac{\kappa p_c}{\kappa\, p_c + p_\ell}, \qquad
    D_{\mathrm{HF}/\ell}(\kappa) \;=\; \frac{p_b + \kappa p_c}{p_b + \kappa p_c + p_\ell},
\end{equation}
for a range of $\kappa$ values.

\begin{figure}[h!]
\centering
\includegraphics[width=0.48\columnwidth]{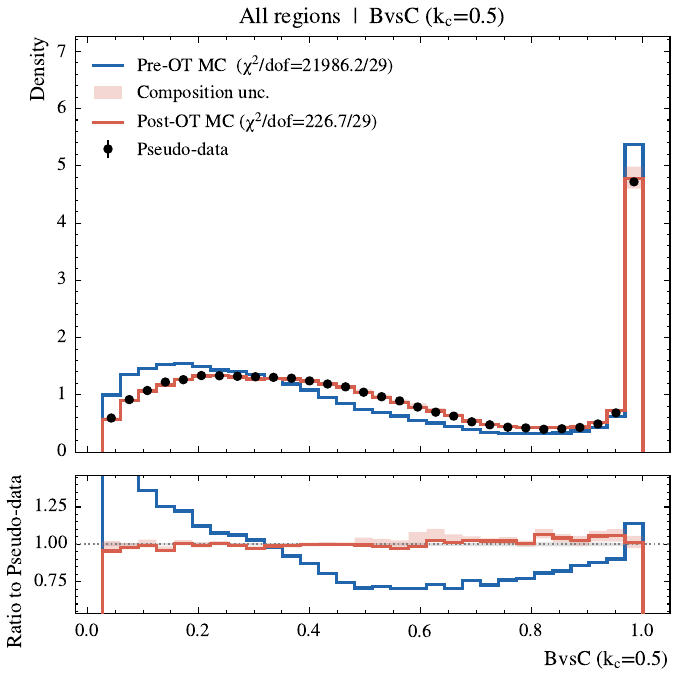}
\hfill
\includegraphics[width=0.48\columnwidth]{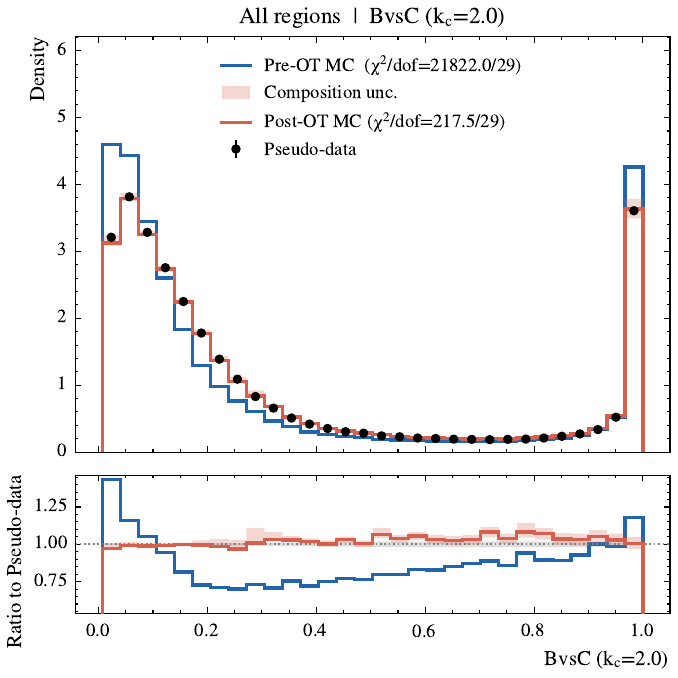}
\caption{Closure of the $b$-vs-$c$ discriminant $D_{b/c}(\kappa)$ in the 1:1:1 validation region for $\kappa=0.5$ (left) and $\kappa=2.0$ (right). Shaded bands indicate composition systematics.}
\label{fig:val_bvsc}
\end{figure}

\begin{figure}[h!]
\centering
\includegraphics[width=0.48\columnwidth]{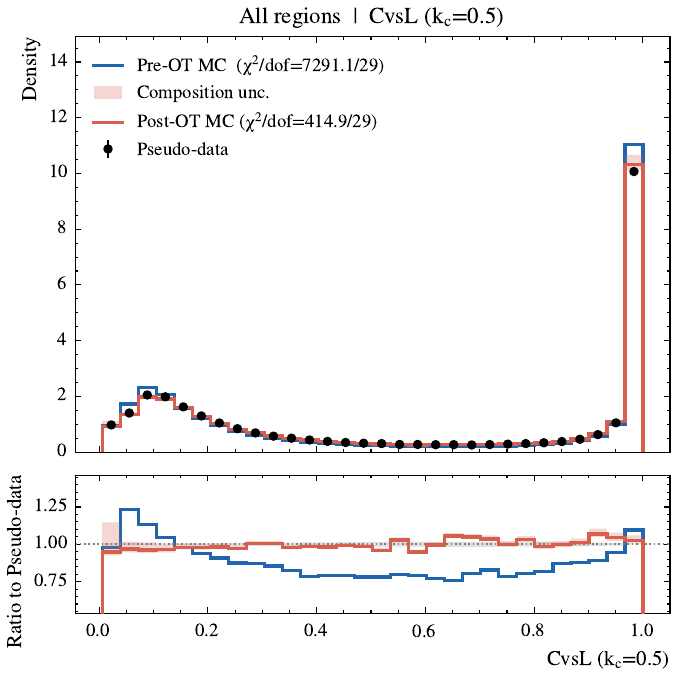}
\hfill
\includegraphics[width=0.48\columnwidth]{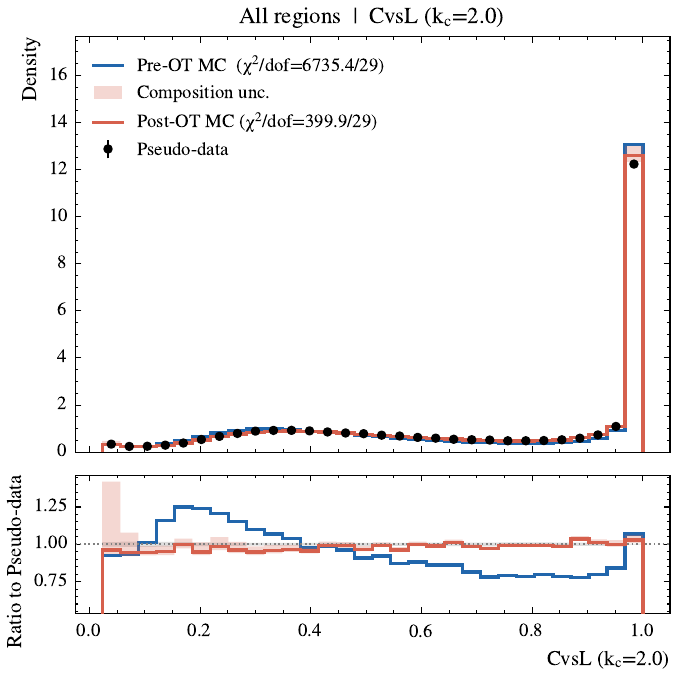}
\caption{Closure of the $c$-vs-light discriminant $D_{c/\ell}(\kappa)$ in the 1:1:1 validation region for $\kappa=0.5$ (left) and $\kappa=2.0$ (right). Shaded bands indicate composition systematics.}
\label{fig:val_cvsl}
\end{figure}

\begin{figure}[h!]
\centering
\includegraphics[width=0.48\columnwidth]{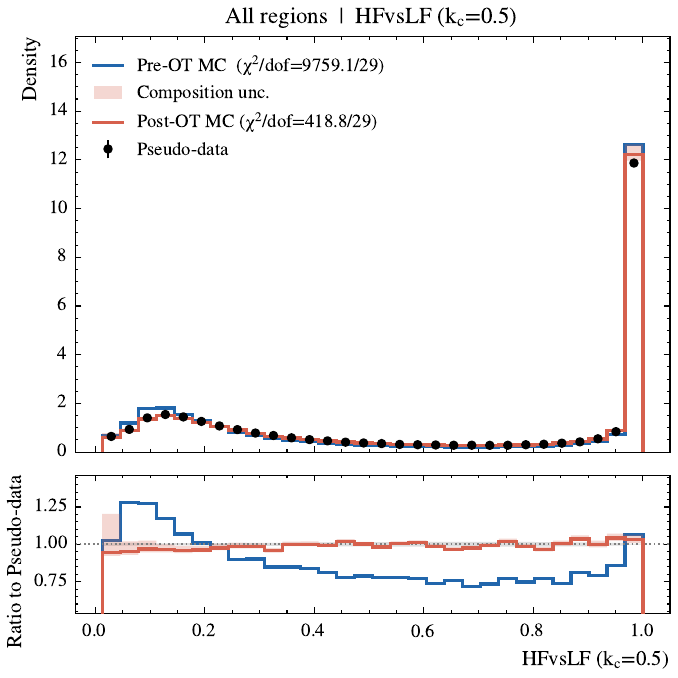}
\hfill
\includegraphics[width=0.48\columnwidth]{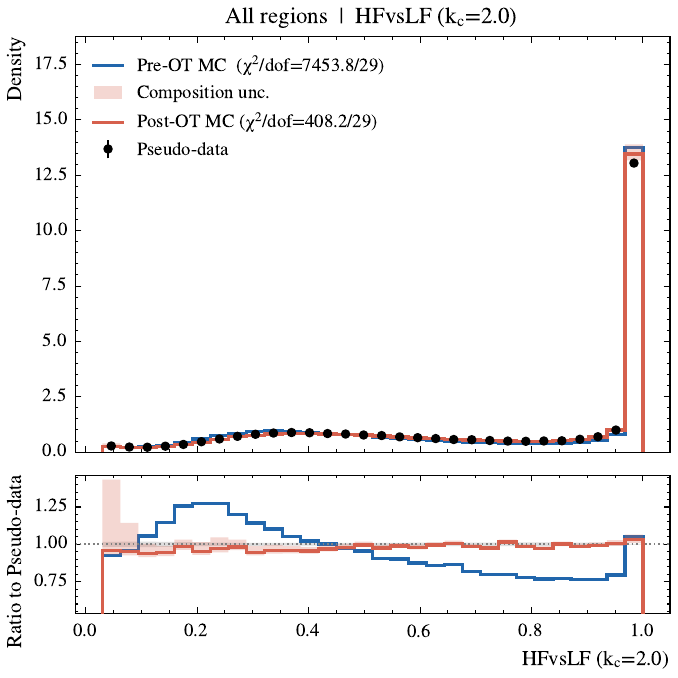}
\caption{Closure of the heavy-flavor-vs-light discriminant $D_{\mathrm{HF}/\ell}(\kappa)$ for $\kappa=0.5$ (left) and $\kappa=2.0$ (right).}
\label{fig:val_hfvslf}
\end{figure}

Figures~\ref{fig:val_bvsc}--\ref{fig:val_hfvslf} show closure for the binary discriminants defined above, evaluated at several representative $\kappa$ values. The $\kappa$ parameter controls the relative weight assigned to the $c$ flavor; varying $\kappa$ corresponds to scanning different working-point operating lines without retraining, and good closure across all $\kappa$ values confirms that a single transport map suffices for the full family of derived discriminants.

Across all discriminants and $\kappa$ values examined, the calibrated distributions show substantially improved agreement with the pseudo-data target relative to the uncorrected MC. The residual mis-closure is generally encompassed by the composition systematic uncertainty bands, indicating that the OT maps trained on the flavor-enriched control regions can transfer to this independent validation mixture without retraining. Within the scope of this pseudo-data study, this supports the view that the flavor-factorized transport is not tied to a single enriched training composition, which is an encouraging feature for downstream applications.

It should be emphasized that, by construction, the control and validation regions in this study are drawn from the same pseudo-data population and differ only in their mixture fractions $\pi_{r,f}$. The flavor-conditional response $q_f(x)$ is therefore identical across regions by definition. Consequently, the present closure tests two specific properties of the proposed framework: (i)~the ability of the EM-extracted targets and the trained transport maps to generalize to mixture compositions not seen during training, and (ii)~the simultaneous validity of a single ILR-space transport map across the family of $\kappa$-parameterized derived discriminants. It does \emph{not} test transferability across physically distinct environments, such as differences in jet kinematics, event topology, or detector conditions, since such differences are absent by construction. Validating that aspect of the framework requires application to real collision data with appropriate phase-space binning, which is left to future experimental studies.
\section{Conclusion}
In this work, we developed a geometry-aware calibration framework for multiclass flavor taggers and studied it in a pseudo-data closure setup, where flavor-conditional targets are extracted from control-region data. The calibration is formulated as a flavor-factorized optimal-transport problem on the probability simplex. By working in the ILR embedding, the quadratic transport cost becomes consistent with the Aitchison geometry of compositional probabilities. To extract flavor-conditional targets directly from control-region data, we combine several enriched regions within an EM-based mixture model with normalizing-flow component densities, and then train Brenier-type transport maps from the MC source distributions to the extracted data targets. In simulation-based closure tests, the method improves closure not only in the enriched control regions used for target extraction, but also in an independent validation mixture, including derived discriminants that are not used explicitly in training.

A main challenge in this program is that the target extraction itself is a high-dimensional, non-convex inference problem. In particular, the simultaneous determination of mixture fractions and flexible component densities can lead to unstable or only weakly constrained directions, so an apparently successful fit should not be taken to imply a unique or fully data-determined solution. To address this, we introduced a practical uncertainty prescription based on the fitted covariance of the mixture composition and its linearized feedback into the extracted component densities. This is not intended as a complete treatment of the covariance of all model parameters. It is, however, a tractable way to identify amplified directions, separate data-constrained modes from those dominated by priors, and assign uncertainty bands to the extracted targets at reasonable computational cost.

More generally, this study shows that machine-learning-based calibration is not automatically stable simply because a flexible model reproduces the observed distributions well. When latent composition, model misspecification, and feedback between nuisance-like quantities and learned densities are all present, the optimization can drift along nearly flat directions or absorb mismodelling in ways that are not obvious from standard closure plots alone. Here we make this point in the specific context of flavor-tagging calibration, but the underlying issue is broader. Calibration procedures based on expressive models require explicit checks of identifiability, stability diagnostics, and some prescription for uncertainty propagation. A related caution applies, in a different way, even to more conventional calibration strategies when the inferred correction depends on imperfectly constrained compositions.
There are several natural directions for future work. One is to incorporate auxiliary inputs into the normalizing-flow component models, so that residual dependence on jet kinematics or the event environment can be modeled directly rather than handled only through phase-space binning. Such a conditional formulation may improve transferability across regions, but it also brings additional challenges in optimization and identifiability that need to be understood. Another is to quantify imperfect transferability itself more explicitly. In the present closure setup, the common flavor-conditional response assumption is imposed by construction, whereas in real data it may be violated by residual differences in kinematics, topology, or detector conditions between control and application regions. Understanding how such violations bias the extracted component densities and the propagated uncertainties is therefore an important next step.

A further direction is to improve the optimization landscape itself, rather than only regularizing the final result. Ideally, one would like parameterizations or training procedures that reduce spurious local minima, limit cross-compensation between mixture fractions and component densities, and make the physically relevant solution more reproducible across different initializations. More structured architectures, stronger inductive biases, or partially constrained update schemes may help, but whether one can design a practically useful setup with substantially better global behavior remains an open question. In this sense, we view the present work not only as a calibration method, but also as a case study illustrating that, for ML-based calibration, geometry, identifiability, transferability, and uncertainty propagation can be just as important as model flexibility itself.


\end{document}